\pdfoutput=1
\RequirePackage{ifpdf}
\ifpdf % We~are running pdfTeX in pdf mode
\documentclass[pdftex]{sigma}
\else
\documentclass{sigma}
\fi

\newcommand{\D}{\mathrm{d}}

\newcommand{\calO}{{\cal O}}

\newcommand{\mo}{{\cal O}}

\newcommand{\mAn}{\mathcal{A}_n}
\newcommand{\mA}{\mathcal{A}}
\newcommand{\an}[1]{\langle#1\rangle}
\newcommand{\sq}[1]{[#1]}
\newcommand{\zb}{\bar{z}}
\newcommand{\rsq}[1]{|#1]}

\newcommand{\lan}[1]{\langle#1|}
\newcommand{\mb}{\bar{m}}

\newcommand{\hb}{\bar{h}}

\usepackage{bm}
\usepackage{mathrsfs}
\usepackage{extarrows}

\numberwithin{equation}{section}

\begin{document}
\allowdisplaybreaks

\newcommand{\arXivNumber}{2111.11356}

\renewcommand{\PaperNumber}{044}

\FirstPageHeading

\ShortArticleName{Deformed $w_{1+\infty}$ Algebras in the Celestial CFT}

\ArticleName{Deformed $\boldsymbol{w_{1+\infty}}$ Algebras in the Celestial CFT}

\Author{Jorge MAGO, Lecheng REN, Akshay YELLESHPUR SRIKANT and Anastasia VOLOVICH}

\AuthorNameForHeading{J.~Mago, L.~Ren, A.~Yelleshpur Srikant and A.~Volovich}

\Address{Department of Physics, Brown University, Providence, RI 02912, USA}
\Email{\href{mailto:jorge_mago@alumni.brown.edu}{jorge\_mago@alumni.brown.edu}, \href{mailto:lecheng_ren@brown.edu}{lecheng\_ren@brown.edu}, \newline \hspace*{13.5mm}\href{mailto:akshay_yelleshjpur_srikant@brown.edu}{akshay\_yelleshpur\_srikant@brown.edu}, \href{mailto:anastasia_volovich@brown.edu}{anastasia\_volovich@brown.edu}}

\ArticleDates{Received January 19, 2023, in final form May 22, 2023; Published online July 04, 2023}

\Abstract{We compute the modification of the $w_{1+\infty}$ algebra of soft graviton, gluon and scalar currents in the celestial CFT due to non-minimal couplings. We find that the Jacobi identity is satisfied only when the spectrum and couplings of the theory obey certain constraints. We comment on the similarities and essential differences of this algebra to~$W_{1+\infty}$.}


\Classification{81T35}

\section{Introduction}
Symmetries have played a central role in shaping our understanding of fundamental physics. They often hint towards simpler or alternate formulations of theories. The Lorentz group ${\rm SO}(3,1)$ of 4D flat spacetime acts as the conformal group ${\rm SL}(2,\mathbb{C})$ on the 2D celestial sphere at null infinity. Any theory living on the celestial sphere naturally inherits this conformal symmetry. S-matrix elements, typically evaluated in a basis of momentum eigenstates, can be rewritten in a basis of boost eigenstates via a Mellin transform (for massless particles)~\cite{Pasterski:2017kqt, Pasterski:2016qvg, Pasterski:2017ylz}. The resulting amplitudes transform as correlation functions of a 2D conformal field theory on the celestial sphere. It is conjectured that a suitable theory on the 2D celestial sphere, called Celestial CFT (CCFT), is the holographic dual to gauge and gravity theories in 4D flat spacetime (see~\cite{Aneesh:2021uzk, Pasterski:2021rjz, Raclariu:2021zjz} for reviews).

Celestial CFTs have additional symmetries beyond the usual 2D conformal symmetry
which follow from leading and subleading soft gluon theorems and leading, subleading and subsubleading soft graviton theorems
(see~\cite{Strominger:2017zoo} for a review).
Remarkably, these symmetries completely fix the leading celestial
operator product expansion (OPE) of conformal primary gluons and gravitons on the celestial sphere in terms of the Euler beta function~\cite{pate2019celestial}.
These OPEs have been also derived in~\cite{Fan:2019emx}
by Mellin transforming the collinear limits of amplitudes.
The Euler beta function has an infinite sequence of poles at integer conformal weights,
associated with conformally soft currents:
the first pole corresponds to the conformally soft current, the second
to the subleading soft current, and
so on.\footnote{Soft theorems in the 4D bulk translate into conformally soft
theorems on the celestial sphere in which the conformal
dimensions of external particles
take on specific integer values~\cite{adamo2019celestial,donnay2019conformally,Fan:2019emx, guevara2019notes, nandan2019celestial, pate2019conformally, puhm2020conformally}.
The infinite tower of soft theorems has been discussed in
\cite{Hamada:2018vrw,Li:2018gnc}.}
The OPEs between the conformally soft positive-helicity gluon and graviton currents
have been obtained in~\cite{Himwich:2021dau} by taking
appropriate residues of the results of~\cite{pate2019celestial}.
These OPEs can then be used to compute the
commutation relations of the modes of the soft currents~\cite{guevara2021holographic}.
Remarkably, it was recently shown by Strominger~\cite{Strominger:2021lvk} that the resulting algebra is the wedge algebra of $w_{1+\infty}$ algebra
which has been extensively studied in the 90s (see~\cite{Bakas:1989xu, BERGSHOEFF1991163,Odake:1990rr,Pope:1991ig, Pope:1989ew, POPE1990191, Pope:1991zka}).
This $w_{1+\infty}$ structure of the symmetry algebra holds for all spins and in the presence of supersymmetry~\cite{Ahn:2021erj, Himwich:2021dau,Jiang:2021ovh}. Additionally, it has also been derived from the
worldsheet perspective~\cite{Adamo:2021lrv, Jiang:2021csc}. The effect of loop corrections and higher derivative interactions on conventional soft theorems has been studied in~\cite{Bern:2014oka,bern1999infrared,bern1998infrared,Cachazo:2014dia, Elvang:2016qvq, he2014loop}. It is interesting to understand their corresponding celestial versions and their implications for the $w_{1+\infty}$ structure.

In this paper, we consider the effect of non-minimal couplings in the bulk on the OPEs of soft currents in the CCFT. We compute the corresponding modifications to the symmetry algebra, and find that while the modified algebra bears some semblance to the well-known $W_{1+\infty}$~\cite{Pope:1989ew}, it differs from $W_{1+\infty}$ in crucial ways. We find that the Jacobi identity imposes constraints on the spectrum and couplings of the theory. The rest of the paper is organized as follows. We review the OPEs of soft currents in the presence of non-minimal couplings in Section~\ref{sec:OPEs}. We compute the algebra in a simplified form and demand the closure of Jacobi identities in Section~\ref{sec:algebra}. In Appendix~\ref{app:singlesoft}, we demonstrate the structure of soft current operators directly from well-known scattering amplitudes.
Appendix~\ref{app:doublesoft} provides an alternate derivation of the gluon soft current OPEs from double soft limits of scattering amplitudes. Finally, in Appendix~\ref{app:altcomm}, we comment on an alternate definition of the commutator for soft currents.

\section{Soft current OPEs}
\label{sec:OPEs}
The OPE for two operators of arbitrary spins, $s_1$, $s_2$ and conformal dimensions $\Delta_1, \Delta_2$ in~CCFT is heavily constrained. They have been determined from collinear limits of scattering amplitudes~\cite{Fan:2019emx} and separately from asymptotic symmetries~\cite{pate2019celestial}. The contribution to the OPE from a~three-point interaction with bulk dimension $p+4$ was worked out in full generality in~\cite{Himwich:2021dau} and takes
the form\footnote{We are only considering the holomorphic collinear pole. Depending on $s_1$, $s_2$, there could be an additional anti-holomorphic pole.}
\begin{align}
 \label{eq:genericOPE}
 \mo_{h_1, \bar{h}_1}(z_1, \zb_1) \mo_{h_2, \bar{h}_2} (z_2, \zb_2)\sim \frac{1}{z_{12}} \sum_{\alpha=0}^\infty C_p^{(\alpha)} \big(\bar{h}_1, \bar{h}_2\big) \zb_{12}^{p+\alpha}\bar{\partial}^\alpha \mo_{h_1+h_2-1, \bar{h}_1+\bar{h}_2+p} (z_2, \zb_2),
\end{align}
where $(h,\hb) = \big( \frac{\Delta+s}{2}, \frac{\Delta-s}{2} \big)$ are the conformal weights. The OPE in (\ref{eq:genericOPE}) gets modified slightly if both operators carry a color index. In this case the right hand side must include an appropriate group theoretical factor. In the rest of this paper, we will avoid writing color indices explicitly unless necessary. The OPE coefficient $ C_p^{(\alpha)} \big(\bar{h}_1, \bar{h}_2\big)$ is completely determined by the three-point amplitude of massless particles with helicities $s_1$, $s_2$, $-s_I$ (where $s_I = s_1 + s_2 - p - 1$ is the spin of the intermediate particle in the collinear factorization channel) to be
\cite{Himwich:2021dau}
\begin{align}
 \label{eq:genericopecoefficient}
 C_p^{(\alpha)} \big(\bar{h}_1, \bar{h}_2\big) &=-\frac{1}{2} \kappa_{s_1, s_2,-s_I} \frac{1}{\alpha!}B\big(2\bar{h}_1 + p + \alpha, 2\bar{h}_2 + p\big),
\end{align}
where $B(a,b)$ is the Euler beta function, $\kappa_{s_1, s_2,-s_I}$ is the coupling constant of the relevant three-point amplitude
which has mass dimension $-p$ which can include contributions from multiple terms in the Lagrangian. As an example, $\kappa_{2,2,2}$ comes from various higher derivative terms with different $R^3$ type contractions. Discounting the existence of massless higher spin particles
in the bulk, there are only a finite number of interactions which contribute to this OPE and we must sum over all allowed values of $p$. It is apparent from (\ref{eq:genericopecoefficient}) that the OPE coefficient has poles in $\Delta$ corresponding to conformally soft limits. The soft current with dimension $k$ for a particle with spin $s$ is defined by the limit
\begin{align*}
 %\label{eq:softcurrentsspin}
 H^{k,s} = \lim_{\Delta \to k} \left(\Delta - k\right) \mo_{\frac{k+s+\epsilon}{2}, \frac{k+\epsilon-s}{2}} = \lim_{\epsilon \to 0} \epsilon \mo_{\frac{k+s+\epsilon}{2}, \frac{k+\epsilon-s}{2}},
\end{align*}
where $k=2,1,0, -1, -2, \dots$ for gravitons, $k = 1, 0 ,-1, \dots$ for gluons and $k=0, -1, -2, \dots$ for scalars.\footnote{Note that the $k=0$ term is non trivial for scalars in the adjoint representation.} Again, it is understood that missing color indices are restored in the appropriate cases. Appendix~\ref{app:singlesoft} contains more details on the derivation and structure of these soft currents. Using this definition of soft currents and the limit
\begin{align*}
 %\label{eq:OPEcoeffarbitraryspins}
 \lim_{\epsilon \to 0} \frac{\epsilon}{2} \frac{\Gamma \big( 2\bar{h}_1 + p + \alpha + \epsilon\big)\Gamma \big(2\bar{h}_2 + p + \epsilon\big)}{\Gamma \big(2\bar{h}_1 + 2\bar{h}_2 + 2p + \alpha + 2\epsilon\big)} = \begin{pmatrix}-2\bar{h}_1-2\bar{h}_2-2p-\alpha\\ -2\bar{h}_2-p\end{pmatrix},
\end{align*}
we obtain the soft current OPE~\cite{Himwich:2021dau}
\begin{align}
 \label{eq:softcurrentOPEallspins}
 &H^{k_1, s_1} (z_1, \zb_1) H^{k_2, s_2} (z_2, \zb_2) \\
 &\sim-\sum_p \frac{ \kappa_{s_1,s_2,-s_I}}{2} \frac{1}{z_{12}}
 \nonumber \sum_{\alpha=0}^{\infty} \frac{(\zb_{12})^{\alpha+p}}{\alpha!} \begin{pmatrix}-2\bar{h}_1-2\bar{h}_2-2p-\alpha\\ -2\bar{h}_2-p\end{pmatrix} \bar{\partial}^\alpha H^{k_1+k_2+p-1,s_1+s_2-p-1}.
\end{align}

The OPE of two positive helicity gravitons includes the term with $p=1$, which is the contribution from the usual MHV amplitude with coupling $\kappa_{-2,2,2}$, $p=3$ which is the contribution from a graviton-graviton-scalar coupling $\kappa_{0,2,2}$, and $p=5$ which is the contribution from a~non-minimal $R^3$ coupling $\kappa_{2,2,2}$. The roster of soft currents considered in this paper is
\begin{itemize}\itemsep=0pt
 \item soft graviton currents $H^{k, \pm 2}$,
 \item soft gluon currents $H^{k, \pm 1, a}$,
 \item two kinds of soft scalar currents $H^{k, 0, a}$ (colored) and $H^{k, 0}$ (uncolored).
\end{itemize}
For the sake of clarity, we explicitly write out two OPEs involving operators of color. The OPE of two positive helicity gluons is
\begin{align}
 \label{eq:softcurrentOPEcolor}
 &H^{k_1, +1, a} (z_1, \zb_1) H^{k_2, +1, b} (z_2, \zb_2)\\
 &\quad \sim-{\rm i}f^{abc}\sum_{p=0,2}\frac{ \kappa_{1,1,p-1}}{2} \frac{1}{z_{12}} \sum_{\alpha} \frac{(\zb_{12})^{\alpha+p}}{\alpha!} \begin{pmatrix}-2\bar{h}_1-2\bar{h}_2-2p-\alpha\\-2\bar{h}_2-p\end{pmatrix} \bar{\partial}^\alpha H^{k_1+k_2+p-1,1-p,c}\nonumber\\
 &\quad\quad -\frac{\delta^{ab}}{N_c}\frac{ \kappa_{1,1,0}}{2} \frac{1}{z_{12}} \sum_{\alpha} \frac{(\zb_{12})^{\alpha+1}}{\alpha!} \begin{pmatrix}-2\bar{h}_1-2\bar{h}_2-2-\alpha\\ -2\bar{h}_2-1\end{pmatrix} \bar{\partial}^\alpha H^{k_1+k_2,0}\nonumber \\
 &\quad\quad -d^{abc} \frac{ \kappa_{1,1,0}}{2} \frac{1}{z_{12}} \sum_{\alpha} \frac{(\zb_{12})^{\alpha+1}}{\alpha!} \begin{pmatrix}-2\bar{h}_1-2\bar{h}_2-2-\alpha\\ -2\bar{h}_2-1\end{pmatrix} \bar{\partial}^\alpha H^{k_1+k_2,0, c} \nonumber \,,
\end{align}
where $f^{abc}$ is the structure of constant and accompanies gluon operators, $\delta^{ab}$ accompanies uncolored scalars, $N_c$ is the number of colors and $d^{abc}$ is a total symmetric tensor defined by $d_{abc} := 2 \operatorname{Tr} \big\lbrack \big\{ T^a , T^b \big\}\, T^c \big\rbrack$ which accompanies colored scalars.

Finally, the OPE of a gluon and a colored scalar is
\begin{align*}
 %\label{eq:softcurrentOPEcolorscalargluon}
 & H^{k_1, +1, a} (z_1, \zb_1) H^{k_2, 0, b} (z_2, \zb_2) \\
 &\nonumber\quad\sim -{\rm i}f^{abc}\frac{ \kappa_{0,0,1}}{2} \frac{1}{z_{12}} \sum_{\alpha} \frac{(\zb_{12})^{\alpha}}{\alpha!} \begin{pmatrix}-2\bar{h}_1-2\bar{h}_2-\alpha\\-2\bar{h}_2\end{pmatrix} \bar{\partial}^\alpha H^{k_1+k_2-1,0,c}\\
 & \quad\quad-d^{abc}\frac{ \kappa_{0,1,1}}{2} \frac{1}{z_{12}} \sum_{\alpha} \frac{(\zb_{12})^{\alpha+1}}{\alpha!} \begin{pmatrix}-2\bar{h}_1-2\bar{h}_2-2-\alpha\\-2\bar{h}_2-1\end{pmatrix} \bar{\partial}^\alpha H^{k_1+k_2,-1, c} \nonumber \,,
\end{align*}
Note that this time $f^{abc}$ accompanies colored scalars while $d^{abc}$ accompanies gluons and that we have no $\delta^{ab}$ term. We will find that the algebra of soft currents involving gravitons and gluons does not satisfy the Jacobi identity unless we include additional scalars (see Section~\ref{sec:jacobi}).

\section{Algebras}
\label{sec:algebra}
The OPEs obtained in the previous sections can be used to compute the commutators of the modes of soft currents. Following the suggestion in Appendix~\ref{app:singlesoft}, we have
\begin{align}
 \label{eq:modeexpansionzbar}
 H^{k, s}(z, \zb) =\overset{-\hb}{\underset{\mb = -\infty}{\sum}}\, \frac{1}{\zb^{\mb+\hb}} H_{\mb}^{k, s}(z).
\end{align}
The modes can also be extracted from the current by the contour integral
\begin{equation*}
 \begin{aligned}
 & H^{k,s}_{\bar{m}}(z) = \underset{|\zb|=\epsilon }{\oint} \frac{\D \bar{z}}{2\pi {\rm i}} \bar{z}^{\bar{m}+\hb-1} H^{k,s}(z , \bar{z}).
 \end{aligned}
\end{equation*}
We define a 2D commutator as
\begin{align}
 \label{eq:2Dcommutator}
 \big[ H_{\mb_1}^{k_1, s_1}, H_{\mb_2}^{k_2, s_2} \big](z_2) =\underset{\mathcal{C}}{\oint} \frac{{\rm d}z_1}{2\pi {\rm i}} \frac{{\rm d}\zb_1}{2\pi {\rm i}} \zb_1^{\mb_1 + \hb_1 - 1} \frac{{\rm d}\zb_2}{2\pi {\rm i}} \zb_2^{\mb_2 + \hb_2 - 1}H^{k_1, s_1}(z_1, \zb_1) H^{k_2, s_2}(z_2, \zb_2),
\end{align}
where the contour $\mathcal{C}$ is defined by the conditions $z_1=z_2$, $|\zb_1|=\epsilon$, $|\zb_2|=\epsilon$ with $\epsilon$ infinitesimal. The OPE of soft currents with opposite spin involves both holomorphic and anti-holomorphic poles. The commutator defined above captures the residue on the holomorphic poles only. We can alternatively define a commutator which is sensitive to both sets of poles. However, this definition would lead to a violation of
the Jacobi identity as we will see
in Appendix~\ref{app:altcomm}. A~particularly striking consequence of the definition (\ref{eq:2Dcommutator}) is
\begin{align*}
 %\label{eq:zerocomm}
 \big[ H_{\mb_1}^{k_1, s_1}, H_{\mb_2}^{k_2, s_2} \big](z_2) = 0 \qquad \text{if} \quad s_1<0 , \ s_2<0,
\end{align*}
since the OPE of two negative helicity soft currents only involves anti-holomorphic poles. The algebra of the modes of the soft currents is determined by the OPE (\ref{eq:softcurrentOPEallspins}) to be
\begin{align*}
 %\label{eq:commutator}
 \big[ H_{\mb_1}^{k_1, s_1}, H_{\mb_2}^{k_2, s_2} \big] &= -\sum_{p} \frac{\kappa_{s_1,s_2,-s_I}}{2} \sum_{\alpha=0}^{\infty} \begin{pmatrix}-2\bar{h}_1-2\bar{h}_2-2p-\alpha\\ -2\bar{h}_2-p\end{pmatrix}\\
 &\nonumber \times\oint \frac{{\rm d}\zb_1}{2\pi {\rm i}} \frac{{\rm d}\zb_2}{2\pi {\rm i}} \zb_1^{\mb_1+\hb_1-1} \zb_2^{\mb_2+\hb_2-1} \frac{(\zb_{12})^{\alpha+p}}{\alpha!} \bar{\partial}_2^\alpha H^{k_1+k_2+p-1, s_1+s_2-p-1} (z_2, \zb_2).
\end{align*}
Performing the $\zb_1$ integral, we get
\begin{align*}
 \nonumber&\big[ H_{\mb_1}^{k_1, s_1}, H_{\mb_2}^{k_2, s_2} \big]=\\
 &\qquad -\sum_{p} \frac{\kappa_{s_1,s_2,-s_I}}{2} \sum_{\alpha=0}^{\infty} \begin{pmatrix}-2\hb_1-2\hb_2-2p-\alpha\\ -2\bar{h}_2-p\end{pmatrix} \binom{p+\alpha}{-\mb_1-\hb_1} \frac{1}{\alpha!}(-1)^{\alpha+p+\mb_1+\hb_1}\\
 &\qquad \times\oint \frac{{\rm d}\zb_2}{2\pi {\rm i}} \zb_2^{p+\alpha+\mb_1+\hb_1+\mb_2+\hb_2-1} \bar{\partial}_2^\alpha H^{k_1+k_2+p-1, s_1+s_2-p-1} (z_2, \zb_2).
\end{align*}
We can perform the last integral by using the mode expansion of $H^{k_1+k_2+p-1, s_1+s_2-p-1} (z_2, \zb_2)$ in (\ref{eq:modeexpansionzbar}). The result is
\begin{align}
 \big[ H_{\mb_1}^{k_1, s_1}, H_{\mb_2}^{k_2, s_2} \big]&= -\sum_{p} \frac{\kappa_{s_1,s_2,-s_I}}{2} \sum_{\alpha=0}^{\infty} \binom{-2\hb_1-2\hb_2-2p-\alpha}{-2\hb_2-p} \binom{p+\alpha}{-\mb_1-\hb_1}\label{eq:HHalgebra1}\\
 & \times\frac{(-1)^{\alpha+p+\mb_1+\hb_1}}{\alpha!} \frac{\bigl(-\mb_1-\mb_2-\hb_1-\hb_2-p\big)!}{\bigl(-\mb_1-\mb_2-\hb_1-\hb_2-p-\alpha\big)!} H_{\mb_1+\mb_2}^{k_1+k_2+p-1, s_1+s_2-p-1}.\nonumber
\end{align}
It is useful to rewrite the above expression using the identity
\begin{equation*}
 \frac{(p+\alpha)!}{\bigl(-\mb_1 - \hb_1\big)! \big(\alpha+p+\mb_1 + \hb_1\big)! \, \alpha!} = \sum_{x=0}^p \binom{p}{x} \frac{1}{\bigl(-\mb_1 - \hb_1 - x\big)! \big( \alpha + x + \mb_1 + \hb_1 \big)!} .
\end{equation*}
The commutation relation (\ref{eq:HHalgebra1}) now becomes
\begin{align*}%\label{eq:HHalgebra2}
\big[ H_{\mb_1}^{k_1, s_1}, H_{\mb_2}^{k_2, s_2} \big]&= -\sum_{p} \frac{\kappa_{s_1,s_2,-s_I}}{2} \sum_{x=0}^{p} \binom{p}{x} \sum_{\alpha=0}^{\infty} \binom{-2\hb_1-2\hb_2-2p-\alpha}{-2\hb_2-p} (-1)^{\alpha + p + \mb_1 + \hb_1}\nonumber \\
 & \times\frac{\bigl( -\mb_1 - \mb_2 - \hb_1 - \hb_2 - p \big)! \,\, H_{\mb_1+\mb_2}^{k_1+k_2+p-1, s_1+s_2-p-1}}{\bigl(-\mb_1 - \mb_2 - \hb_1 - \hb_2 - p - \alpha\big)! \bigl( -\mb_1 - \hb_1 - x \big)! \big( \alpha + x + \mb_1 + \hb_1 \big)!}.
\end{align*}
This allows us to perform the sum over $\alpha$ more easily to get
\begin{align}
 \label{eq:general}
 \big[ H_{\mb_1}^{k_1, s_1}, H_{\mb_2}^{k_2, s_2} \big]
 = -\sum_{p} \sum_{x=0}^p &\left[\frac{\kappa_{s_1,s_2,-s_I}}{2}
 (-1)^{p-x}\binom{p}{x} \binom{\mb_1+\mb_2-\hb_1-\hb_2-p}{\mb_1-\hb_1-p+x} \right.\\
 & \quad \times\left.\binom{-\mb_1-\mb_2-\hb_1-\hb_2-p}{-\mb_1-\hb_1-x} \right]
 H^{k_1 + k_2 + p - 1, s_1 + s_2 -p-1}_{\bar{m}_1 + \bar{m}_2}. \nonumber
\end{align}
Restricting the sum over $p$ to only include particles with helicities between $-2$ and $2$ implies that $p \in \left( \operatorname{Max} (s_1+s_2-3,0), \,\operatorname{Max} (s_1+s_2+1, 0)\right)$. However, some values of $p$ in this range might still be forbidden if the corresponding amplitude does not exist. The introduction of color indices modifies this formula in a manner analogous to the modification of the OPE.

\subsection[W\_infty-like currents]{$\boldsymbol{W_\infty}$-like currents}
%\label{sec:winf-like}
It is useful to define new currents which are light-transforms of the soft currents. These have simplified the algebra and revealed its $w_\infty$ structure in~\cite{Himwich:2021dau, Jiang:2021ovh, Strominger:2021lvk}. Let\footnote{As usual color indices are suppressed.}
\begin{align*}
% \label{eq:winf}
 W^{q,s} (z, \zb) = \Gamma(2q) {\bf {\bar{L}}}\big[ H^{s+2(1-q),s}(z, \zb) \big],
\end{align*}
{where $q = 1, \frac{3}{2}, 2, \dots$ and} %where $q = 1, \frac{3}{2}, 2, \dots$ for particles with helicity $s>0$, $q = s+1, s+3/2, s+2, \dots $ for particles with helicity $s<0$ and
\begin{align*}
 %\label{eq:lighttransform}
 {\bf \bar{L}}[\mo_{h,\hb} (z, \zb)] = \int_{\mathbb{R}} \frac{{\rm d}\bar{w}}{(\zb-\bar{w})^{2-2\hb}} \mo_{h,\hb} (z, \bar{w}).
\end{align*}
The conformal weights of $W^{q,s}$ are $(s+1-q, 1-q)$. Applying this definition to the mode expansion (\ref{eq:modeexpansionzbar}), we get
\begin{align*}
 %\label{eq:lighttransformsoftplus}
 &W^{q,s}(z,\bar{z}) = \sum_{\mb=-\infty}^{\mb=-\hb}(-\mb+q-1)! (\mb+q-1)!\frac{1}{\zb^{\mb+q}} H_{\mb}^{s+2(1-q),s} (z),
\end{align*}
which implies the following mode expansion
\begin{align*}
 W^{q,s}(z,\bar{z}) = \sum_{\mb=-\infty}^{q-1} W^{q,s}_{\mb}(z) \frac{1}{\zb^{\mb+q}}
\end{align*}
with\footnote{The expression as written is valid only for modes in the range $1-q\leq \mb \leq q-1$. Similar expressions can be written for modes outside this range.}
\begin{align*}
W^{q,s}_{\mb} (z)= (-\mb+q-1)! (\mb+q-1)!H_{\mb}^{s+2(1-q),s} (z).
\end{align*}
The commutator (\ref{eq:general}) takes the simplified form
\begin{align}
 \nonumber
 \big\lbrack W_{\bar{m}_1}^{q_1,s_1} ,\, W_{\bar{m}_2}^{q_2,s_2} \big\rbrack = -\sum_{p}\sum_{x=0}^p &\frac{\kappa_{s_1,s_2,-s_I}}{2} (-1)^{p-x} \binom{p}{x}\, \lbrack \bar{m}_1 + q_1 - 1 \rbrack_{p-x} \lbrack -\bar{m}_1 + q_1 - 1 \rbrack_{x} \\
 \times\, & \lbrack \bar{m}_2 + q_2 - 1 \rbrack_{x} \lbrack -\bar{m}_2 + q_2 - 1 \rbrack_{p-x} \ W_{\bar{m}_1 + \bar{m}_2}^{q_1+q_2-p-1, s_1+s_2-p-1}\nonumber\\
&\hspace{-3.7cm}\equiv-\sum_{p=\operatorname{Max}(s_1+s_2-3,0)}^{\operatorname{Max}(s_1+s_2+1,0)} \frac{\kappa_{s_1,s_2,-s_I}}{2} N(q_1, q_2, \mb_1, \mb_2, p)
 W_{\bar{m}_1 + \bar{m}_2}^{q_1+q_2-p-1, s_1+s_2-p-1},\label{eq:Wgeneral}
\end{align}
where $\lbrack a \rbrack_n := a (a-1) \cdots (a-n+1)$ is the descending Pochhammer symbol and
\begin{gather}
 \label{eq:Ndef}
 N(q_1, q_2, \mb_1, \mb_2, p) = \sum_{x=0}^p (-1)^{p-x} \binom{p}{x}\, \lbrack \bar{m}_1 + q_1 - 1 \rbrack_{p-x} \lbrack -\bar{m}_1 + q_1 - 1 \rbrack_{x} \nonumber\\
 \hphantom{N(q_1, q_2, \mb_1, \mb_2, p) =}{} \times \lbrack \bar{m}_2 + q_2 - 1 \rbrack_{x} \lbrack -\bar{m}_2 + q_2 - 1 \rbrack_{p-x}.
\end{gather}
While $p$ generically ranges between $\operatorname{Max}(s_1+s_2-3,0)$ and $\operatorname{Max}(s_1+s_2+1,0)$, some values of $p$ are forbidden because the corresponding amplitudes do not exist. Note that $ N(q_1, q_2, \mb_1, \mb_2, p)$ is symmetric under the exchange of $q_1 \leftrightarrow q_2$, $\mb_1 \leftrightarrow \mb_2$ for even $p$ while it is antisymmetric for odd~$p$. Thus, when both $W_{\mb_1}^{q_1,s_1} $ and $W_{\mb_2}^{q_2,s_2}$ have no color indices, the commutator is antisymmetric under the exchange only for odd values of $p$. When we include colored particles, the allowed values of $p$ depend on the group theoretical structure. For example, the commutator involving two gluons then reads (note that $N\left(q_1, q_2, \mb_1, \mb_2, 0\right) = 1)$
\begin{align*}
 %\label{eq:Wallgluons}
 &\big\lbrack W_{\bar{m}_1}^{q_1,s_1,a} ,\, W_{\bar{m}_2}^{q_2,s_2,b}\big\rbrack \\
&\qquad =
 -{\rm i}f^{abc} \sum_{\substack{p=\operatorname{Max}(s_1+s_2-2,0) \\ p \text{ even}}}^{\operatorname{Max}(s_1+s_2,0)} \frac{\kappa_{s_1, s_2, -s_I}}{2} N(q_1, q_2, \mb_1, \mb_2, p ) W_{\bar{m}_1 + \bar{m}_2}^{q_1+q_2-1, s_1+s_2-p-1,c} \nonumber\\
 &\qquad\quad -\frac{\delta^{ab}}{N_c}\sum_{\substack{p=\operatorname{Max}(s_1+s_2-3,0)\\p \text{ odd} }}^{\operatorname{Max}(s_1+s_2+1,0)} \frac{\kappa_{s_1,s_2,-s_I}}{2} N(q_1, q_2, \mb_1, \mb_2, p) W_{\bar{m}_1 + \bar{m}_2}^{q_1+q_2-p-1, s_1+s_2-p-1} \nonumber \\
 &\qquad\quad -d^{abc}\sum_{\substack{p=\operatorname{Max}(s_1+s_2-2,0)\\p \text{ odd}}}^{\operatorname{Max}(s_1+s_2,0)} \frac{\kappa_{s_1,s_2,-s_I}}{2} N(q_1, q_2, \mb_1, \mb_2, p) W_{\bar{m}_1 + \bar{m}_2}^{q_1+q_2-p-1, s_1+s_2-p-1, c} \nonumber,
\end{align*}
while the commutator of a gluon (or colored scalar) with a graviton or scalar reads
\begin{align*}
 %\label{eq:Wgluongraviton}
 &\big\lbrack W_{\bar{m}_1}^{q_1,s_1,a} ,\, W_{\bar{m}_2}^{q_2,s_2}\big\rbrack \\
&\qquad =-\sum_{\substack{p=\operatorname{Max}(s_1+s_2-2,0)\\p \text{ odd}}}^{\operatorname{Max}(s_1+s_2,0)} \frac{\kappa_{s_1,s_2,-s_I}}{2} N(q_1, q_2, \mb_1, \mb_2, p) W_{\bar{m}_1 + \bar{m}_2}^{q_1+q_2-p-1, s_1+s_2-p-1, a}. \nonumber
\end{align*}

\subsection{Jacobi identity}
\label{sec:jacobi}
We must now demand that these commutators satisfy the Jacobi identity. To this end, we compute the double commutator from (\ref{eq:Wgeneral}). After some algebra, we have
\begin{align}
 \nonumber
 &\big[ \big[ W_{\mb_1}^{q_1,s_1}, W_{\mb_2}^{q_2,s_2} \big],W_{\mb_3}^{q_3,s_3}\big] + \text{cyclic} \\
& \qquad=\nonumber \sum_{\substack{l =s_1+s_2+s_3-4\\ l \text{ even}}}^{s_1+s_2+s_3} \sum_{\substack{r = 0\\ r \text{ odd}}}^l \frac{\kappa_{s_1, s_2, 1+l-r-s_1-s_2}}{2}
 \frac{\kappa_{s_1+s_2-l-1+r, s_3, l+2-s_1-s_2-s_3}}{2} \\
 &\nonumber \qquad\qquad\qquad\quad\qquad \times N(q_1, q_2, \mb_1, \mb_2, l-r) N(q_1+q_2-l-1+r, q_3, \mb_1+\mb_2, \mb_3,r) \\
 & \qquad\qquad\qquad\quad \qquad\times W^{q_1+q_2+q_3-l-2, s_1+s_2+s_3-l-2}_{\mb_1+\mb_2+\mb_3} + \text{cyclic}.\label{eq:posgravjoacobi}
\end{align}
We find that the coupling constants are highly constrained by the Jacobi identities.\footnote{After the first version of this paper was published,~\cite{Ball:2022bgg} showed that in order for the Jacobi identity to be satisfied, only the soft generators inside the wedge $\hb \leq \mb \leq -\hb$ are allowed. } The Jacobi identity for $s_1 = s_2 = s_3 = 2$ requires
\begin{align}
 \label{eq:constraint1}
 (\kappa_{-2,2,2} - \kappa_{0,0,2})\, \kappa_{0,2,2} = 0, \qquad 3\kappa_{0,2,2}^2 = 10 \kappa_{-2,2,2}\, \kappa_{2,2,2}
\end{align}
%The Jacobi identity involving just gravitons fails to close unless we include a scalar (uncolored) soft current and an $R^3$ interaction, both with uniquely determined coefficients.
and the one for $s_1 = s_2 = 2$, $s_3 = 0$ also requires
\begin{equation*}
 (\kappa_{-2,2,2} - \kappa_{0,0,2})\, \kappa_{0,0,2} = 0 .
\end{equation*}

Adding on the gluons, the Jacobi identity for $s_1 = s_2 = 2$, $s_3 = 1$ requires
\begin{equation*}
 (\kappa_{-2,2,2} - \kappa_{-1,1,2})\,\kappa_{-1,1,2} = 0, \qquad \kappa_{-2,2,2}\,\kappa_{1,1,2} = \kappa_{-1,1,2}\,\kappa_{1,1,2} = \kappa_{0,2,2}\,\kappa_{0,1,1} \,,
\end{equation*}
the one for $s_1 = 2$, $s_2 = 1$, $s_3 = 0$ requires
\begin{equation*}
 (\kappa_{0,0,2} - \kappa_{-1,1,2}) \, \kappa_{0,1,1} = 0 ,
\end{equation*}
the one for $s_1 = 2$, $s_2 = s_3 = 1$ requires
\begin{align*}
 %\label{eq:constraint2}
 \kappa_{-1,1,2}\,\kappa_{1,1,1} = 3\kappa_{1,1,2}\,\kappa_{-1,1,1} ,
 %&\frac{\kappa_{0,1,1}}{\kappa_{-2,2,2} } = \frac{\kappa_{1,1,2}}{\kappa_{0,2,2} } \qquad \frac{\kappa_{-1,1,1}}{\kappa_{-1,1,2}} = \frac{\kappa_{1,1,1}}{3 \kappa_{1,1,2}}
\end{align*}
the one for $s_1 = +2$, $s_2 = 1$, $s_3 = -1$ requires
\begin{equation*}
 (\kappa_{-1,1,2} - \kappa_{-2,2,2})\,\kappa_{-1,1,2} = 0 ,
\end{equation*}
the one for $s_1 = s_2 = 1$, $s_3 = 0$ requires
\begin{equation*}
 (\kappa_{-1,1,1} - \kappa_{0,0,1})\, \kappa_{0,0,1} = 0, \qquad (\kappa_{-1,1,1} - \kappa_{0,0,1})\, \kappa_{0,1,1} = 0 ,
\end{equation*}
 and finally, the one for $s_1 = s_2 = s_3 = 1$ requires
\begin{equation*}\label{eq:constraint3}
 \kappa_{0,1,1}^2 = 2\kappa_{-1,1,1}\, \kappa_{1,1,1} .
\end{equation*}
For the final case, the Jacobi identity involving just $F^3$ correction $\kappa_{1,1,1}\neq 0$ fails to close unless we include a colored scalar soft current {\it in addition} to the uncolored scalar, with the coefficient the same as that of uncolored scalar.

\subsection[Relation to W\_\{1+infty\}]{Relation to $\boldsymbol{W_{1+\infty}}$}
The algebra in (\ref{eq:Wgeneral}) is a deformation of the $w_{1+\infty}$ algebra obtained in~\cite{Strominger:2021lvk} in the absence of higher derivative corrections.\footnote{There are more generators in the algebra (\ref{eq:Wgeneral}) than in $w_{1+\infty}$.} There, it was conjectured that quantization of the theory would deform $w_{1+\infty}$ to the well studied
$W_{1 + \infty}$~\cite{Bakas:1989xu, BERGSHOEFF1991163, Fairlie:1990wv, Odake:1990rr,Pope:1991ig,Pope:1989ew,POPE1990191, Pope:1991zka}. This algebra was derived in~\cite{Odake:1990rr,Pope:1989ew} by starting with a generic deformation { with parameter $\lambda$} of $w_{1+\infty}$
\begin{equation*}%\label{eq:WV}
 \big\lbrack V^{q_1}_{\bar{m}_1} ,\, V^{q_2}_{\bar{m}_2} \big\rbrack := \sum_{r=0}^{\infty} \lambda^{2r} g_{2r}^{q_1 q_2}(\bar{m}_1, \bar{m}_2) V_{\bar{m}_1 + \bar{m}_2}^{q_1 + q_2 - 2r} + \lambda^{2q_1} c_{q_1}(\bar{m}) \delta^{q_1 , q_2} \delta_{\bar{m}_1 + \bar{m}_2 , 0},
\end{equation*}
(here $\bar{m}=0,1,2, \dots $ label the modes of $V$) and demanding that the central charge be non-vanishing, the structure constants satisfy the Jacobi identity while also enforcing the truncation of this algebra to a finite number of terms. In particular,
$g_{2r}^{i\,j}(\mb_1, \mb_2,q_1, q_2) =0$ whenever $q_1+q_2-2r <0$. These conditions completely fix the central charge as well as the structure constants to
\begin{gather}
 c_{q} = \frac{2^{2q-3} q! (q+2)!}{(2q+1)!! (2q+3)!!} c, \nonumber\\
 g_{2r}^{q_1 q_2}(\bar{m}_1, \bar{m}_2) = \frac{\phi_{2r}^{q_1 q_2}}{2 (2r+1)!} N(q_1+2, q_2+2, \bar{m}_1 , \bar{m}_2, 2r+1),
\label{eq:Winfstructureconstants}
\end{gather}
with
\begin{equation*}
 \phi_{2r}^{q_1 q_2} = \sum_{a=0}^r \prod_{l=1}^a \frac{(2l-3)(2l+1)(2r-2l+3)(r-l+1)}{l(2 q_1 - 2l + 3)(2 q_2 - 2l + 3)(2 q_1 + 2 q_2 - 4r + 2l + 3)}.
\end{equation*}
Note that $N$ in (\ref{eq:Winfstructureconstants}) is identical to the one appearing in (\ref{eq:Ndef}). Note that when $\lambda = 0$ the algebra goes back to $w_{1+\infty}$ algebra. The algebra obtained in this paper is similar to the algebra referred to as case (1) in~\cite{Fairlie:1990wv}. The function $ \phi_{2r}^{q_1 q_2}$ which has non-trivial zeroes enforcing the truncation of the algebra is absent here. Here the truncation occurred naturally due to the absence of massless higher spin fields.\footnote{During the preparation of the second version of this paper,~\cite{Bu:2022iak} appeared with more detailed studies on this deformed algebra. Paper~\cite{Bu:2022iak} related the algebra in (\ref{eq:Wgeneral}) with the loop algebra of a wedged-symplecton algebra, denoted by $LW_{\wedge}$. The symplecton algebra is a different deformation of $w_{1+\infty}$ from $W_{1+\infty}$ with exactly $\phi_{2r}^{q_1 q_2} = 1$ as we have in (\ref{eq:Wgeneral}).}

It is important to understand how the algebra in (\ref{eq:Wgeneral}) is further deformed by loop corrections. It seems plausible that they would lead to a central extension of the algebra. The scalar currents are forced to exist for the algebra to close in the presence of non-minimal couplings. Clarifying their origin, perhaps by linking them to the spontaneous breaking of some symmetry would also be of interest. Finally, the constraints on the couplings imposed by the Jacobi identity are very restrictive. We leave a systematic study of the implications of these constraints to future work.

\subsection{Simplest loop corrections and connection with the self-dual sector}
One-loop corrections to gluon amplitudes are generically logarithmic and IR divergent. This poses a hurdle to defining a mode expansion of the soft factor. However, one-loop amplitudes involving only positive helicity gluons or only positive helicity gravitons\footnote{These amplitudes vanish at tree-level.} are purely algebraic and their effect on the algebra is easily determined. The one-loop all plus gluon amplitude
is~\cite{bern1994one,mahlon1994multigluon}
\begin{equation*}
 A_{n ; 1}^{[1]}\big({1}^{+}, {2}^{+}, \ldots, {n}^{+}\big)=-\frac{{\rm i}}{48 \pi^{2}} \sum_{1\leq i < j < k < l \leq n} \frac{[i j]\langle j k\rangle[k l]\langle l i\rangle}{\langle 12\rangle\langle 23\rangle {\cdots} \langle n 1\rangle}.
\end{equation*}
A similar expression can be obtained for the corresponding one-loop graviton amplitude~\cite{Bern:1998xc}. Remarkably, the OPEs and algebra computed from the collinear limits of this amplitude are identical to the ones obtained from tree-level amplitudes. They leave the $w_{1+\infty}$ algebra undeformed.

This one-loop amplitude also happens to be the only non-zero amplitude in the self-dual sector of Yang--Mills theory. More precisely, the only non-zero amplitudes in self-dual Yang--Mills and self-dual gravity are the three-point tree-level amplitudes with two-positive and one-negative helicities, and the one-loop amplitudes with all positive helicities~\cite{Chalmers:1996rq}. Consequently, we can only define collinear limits of positive helicity particles in the self-dual sector and the OPEs obtained from them agree with the OPEs of positive helicity particles obtained from the full theory. It would be interesting to analyze the effect of non-minimal couplings (obeying the self-duality condition) on the self-dual sector. Such a theory will naturally be free of anti-holomorphic poles and provides an alternate, and arguably more natural way to deform the~$w_{1+\infty}$ algebra.

\appendix

\section{Single soft factors}
\label{app:singlesoft}
We start by recalling that the Mellin transform of the amplitude is interpreted as a correlation function of gluon primary operators~\cite{Pasterski:2016qvg,Pasterski:2017ylz}. More specifically, we first decompose the amplitude in trace basis as
\begin{align}
 \label{eq:colorordered}
 A_n(1, 2, 3, \dots, n) = \sum_{\sigma \in S_{n-1}} \operatorname{Tr}[T^{a_1} T^{a_{\sigma_2}} \dots T^{a_{\sigma_{n-1}}} ] \mAn (1, 2, 3, \dots, n).
\end{align}
Amplitudes are usually expressed as a function of spinor helicity variables $p_{\alpha \dot{\alpha}} = \lambda_\alpha \tilde{\lambda}_{\dot{\alpha}}$. We parametrize these as
\begin{align}
\label{eq:vars}
 \lambda^\alpha_i = \varepsilon_i \sqrt{\omega_i}\, t_i\begin{pmatrix} 1\\ z_i \end{pmatrix} \qquad \tilde{\lambda}_i^{\dot{\alpha}} = \sqrt{\omega_i}\, t_i^{-1} \begin{pmatrix} 1 \\ \bar{z}_i \end{pmatrix}
\end{align}
and Mellin transform the color-ordered amplitudes,
\begin{align*}
 %\label{eq:++softcelestdef}
 \tilde{\mAn} (1, 2, 3, \dots, n) &:=\int {\rm d}\omega_1 \cdots {\rm d}\omega_n \,\omega_1^{\Delta_1-1} \cdots \omega_n^{\Delta_n-1}\mAn (1, 2, 3, \dots, n).
\end{align*}
Then we have
\begin{align*}
 \operatorname{Tr}[T^{a_1} T^{a_2} \dots T^{a_n} ] \tilde{\mAn} (1, 2, 3, \dots, n)=\big\langle \mo^{a_1}_{\Delta_1}(z_1, \zb_1) \mo^{a_1}_{\Delta_2}(z_2, \zb_2) \cdots \mo^{a_n}_{\Delta_n} (z_n, \zb_n)\big\rangle.
\end{align*}
When particle 1 becomes soft, we can expand the amplitude in a Laurent series as
\begin{align*}
 %\label{eq:powerexpamp}
 \mAn (1, 2, \dots, n) = \sum_{r=-\infty}^{1} \omega_1^{-r} \mAn'^{(r)},
\end{align*}
where the coefficients $\mAn'^{(r)}$ are independent of $\omega_1$. The conformally soft limit~\cite{donnay2019conformally, pate2019conformally} $\Delta_1 \to k$, with $k=1,0, -1, \dots $ gives
\begin{align}
\nonumber
 &\lim_{\Delta_1 \to k} (\Delta_1-k)\big\langle\mo^{a_1}_{\Delta_1}(z_1, \zb_1) \mo^{a_2}_{\Delta_2}(z_2, \zb_2) \dots \mo^{a_n}_{\Delta_n} (z_n, \zb_n)\big\rangle \\
 \nonumber &\qquad=\,\operatorname{Tr}\,[T^{a_1} T^{a_2} \dots T^{a_n} ] \int \frac{\D\omega_1}{\omega_1} (\Delta_1-k)\, \omega_1^{\Delta_1} \sum_{r=-\infty}^{1} \omega_1^{-r} \int \frac{\D\omega_2}{\omega_2} \cdots \frac{\D\omega_n}{\omega_n} \, \omega_2^{\Delta_2} \cdots \omega_n^{\Delta_n} \mAn'^{(r)} \\
 &\qquad=\,\operatorname{Tr}\,[T^{a_1} T^{a_2} \dots T^{a_n} ] \int \frac{\D\omega_2}{\omega_2} \cdots \frac{\D\omega_n}{\omega_n} \, \omega_2^{\Delta_2} \cdots \omega_n^{\Delta_n} \mAn'^{(k)},\label{eq:softcurrentcorrelator}
\end{align}
where from the last but one line to the last line we are using the identity
\begin{equation}
\label{deltaidentity}
\lim_{\epsilon \to 0} \epsilon\omega^{\epsilon-1} = \delta(\omega).
\end{equation}

We interpret this soft expansion as defining the correlation function of soft currents, i.e.,
\begin{align}
 \label{eq:softcurrentdef}
 &\lim_{\Delta_1 \to k} (\Delta_1-k)\big\langle\mo^{a_1,+1}_{\Delta_1}(z_1, \zb_1) \mo^{a_2}_{\Delta_2}(z_2, \zb_2) \cdots \mo^{a_n}_{\Delta_n} (z_n, \zb_n)\big\rangle \nonumber\\
 &\qquad = \big\langle H^{k,+1,a_1} (z_1, \zb_1) \mo^{a_2}_{\Delta_2}(z_2, \zb_2) \cdots \mo^{a_n}_{\Delta_n} (z_n, \zb_n)\big\rangle,\nonumber\\
 &\lim_{\Delta_1 \to k} (\Delta_1-k)\big\langle\mo^{a_1,-1}_{\Delta_1}(z_1, \zb_1) \mo^{a_2}_{\Delta_2}(z_2, \zb_2) \cdots \mo^{a_n}_{\Delta_n} (z_n, \zb_n)\big\rangle \nonumber\\
 &\qquad = \big\langle H^{k,-1,a_1} (z_1, \zb_1) \mo^{a_2}_{\Delta_2}(z_2, \zb_2) \cdots \mo^{a_n}_{\Delta_n} (z_n, \zb_n)\big\rangle,
\end{align}
where $k=1,0, -1, \dots$. All the dependence of the correlation function on the variables $z_1$, $\zb_1$ is entirely due to the soft current operator. We can thus attempt to guess the structure of the soft current operators $H^{k,+1,a}$ by examining the functional dependence of the amplitudes on $z_1$, $\zb_1$ in the soft limit. We first review the case of MHV amplitudes and see the polynomial structure of the currents. We will then move beyond MHV amplitudes and consider the structure of soft expansion of the 6-point NMHV amplitude and non-minimally coupled amplitudes.
\subsection{MHV amplitudes}
We are interested in the structure of amplitudes in the soft limit. The soft limits of negative helicity particles vanish in the MHV sector. Consider the following expression for $\mAn^{\text{MHV}}(1, \dots, n)$, obtained from a BCFW expansion using the shifts $\hat{\lambda}_1 = \lambda_1 + z \lambda_n$, $\hat{\tilde{\lambda}}_n = \tilde{\lambda}_n - z \tilde{\lambda}_1$,
\begin{align}
\label{eq:MHVfactorization}
 \mAn^{\text{MHV}} = \frac{\an{n2}}{\an{n1}\an{12}} \exp \Bigg( \frac{\an{n1}}{\an{n2}} \tilde{\lambda}_1 \frac{\partial}{\partial \tilde{\lambda}_2} + \frac{\an{12}}{\an{n2}} \tilde{\lambda}_1 \frac{\partial}{\partial \tilde{\lambda}_n} \Bigg) \mA_{n-1}^{\text{MHV}} \left(2, \dots, n\right).
\end{align}
Expressing this in terms of $\omega_i$, $z_i$, $\zb_i$, $t_i$ using
\begin{align*}
 %\label{eq:derivatives}
\frac{\partial}{\partial \tilde{\lambda}_i^{\dot{1}}} = \frac{1}{\sqrt{\omega_i}} \left( \omega_i \frac{\partial}{\partial \omega_i} - \bar{z}_i \frac{\partial}{\partial \bar{z}_i} - t_i \frac{\partial}{\partial t_i} \right), \qquad \frac{\partial}{\partial \tilde{\lambda}_i^{\dot{2}}} = \frac{t}{\sqrt{\omega}} \frac{\partial}{\partial \bar{z}_i}, \qquad t_i \frac{\partial}{\partial t_i} \tilde{\mA}_{n-1} = -2 s_i \tilde{\mA}_{n-1},
\end{align*}
gives
\begin{align*}
 %\label{eq:collinearcelestialvars}
 \mA_n = -\frac{1}{\omega_1}\frac{z_{n2}}{z_{n1}z_{12}} \exp & \Bigg\lbrack\frac{z_{n1}}{z_{n2}}\frac{\omega_1}{\omega_2}\left(\omega_2\frac{\partial}{\partial\omega_2}+ \bar{z}_{12} \frac{\partial}{\partial \bar{z}_2} - t_2 \frac{\partial}{\partial t_2}\right)\\
 \nonumber &\quad+ \frac{z_{12}}{z_{1n}}\frac{\omega_1}{\omega_n}\left(\omega_n\frac{\partial}{\partial\omega_n}+ \bar{z}_{1n} \frac{\partial}{\partial \bar{z}_n} - t_n \frac{\partial}{\partial t_n}\right) \Bigg\rbrack {\mA}_{n-1} \left(2, \dots n\right)\\
 \nonumber = -\frac{1}{\omega_1}\frac{z_{n2}}{z_{n1}z_{12}} \exp & \Bigg\{\omega_1\left\lbrack \left(\frac{z_{n1}}{z_{n2}}\frac{\partial}{\partial\omega_2}+ \frac{z_{12}}{z_{n2}}\frac{\partial}{\partial\omega_n}\right) + \left(\frac{z_{n1}\bar{z}_{12}}{z_{n2}\omega_2} \frac{\partial}{\partial \bar{z}_2} + \frac{z_{12}\bar{z}_{1n}}{z_{n2}\omega_2} \frac{\partial}{\partial \bar{z}_n}\right) \right.\\
 \nonumber &\quad\left.+ \left( \frac{z_{n1}}{z_{n2}}\frac{2s_2}{\omega_2} + \frac{z_{12}}{z_{n2}}\frac{2s_n}{\omega_n} \right) \right\rbrack\Bigg\}\mA_{n-1}.
\end{align*}
Using the definition in (\ref{eq:softcurrentdef}) and the relation (\ref{eq:softcurrentcorrelator}), the correlation function of the soft current~$H^{k_1,+1,a_1}$ is
\begin{align*}
%\label{eq:MHVsoftcorrelator}
 &\big\langle H^{k_1,+1,a_1} (z_1, \zb_1) \mo^{a_2}_{\Delta_2}(z_2, \zb_2) \cdots \mo^{a_n}_{\Delta_n} (z_n, \zb_n)\big\rangle \\
& \nonumber =\operatorname{Tr}\,[T^{a_1} T^{a_2} \dots T^{a_n} ] \int \frac{\D\omega_2}{\omega_2} \cdots \frac{\D\omega_n}{\omega_n} \, \omega_2^{\Delta_2} \cdots \omega_n^{\Delta_n} \mAn'^{(k)} \frac{z_{n2}}{z_{n1}z_{12}}\\
& \nonumber = -\operatorname{Tr}\,[T^{a_1} T^{a_2} \dots T^{a_n} ] \ \int \frac{\D\omega_2}{\omega_2} \cdots \frac{\D\omega_n}{\omega_n} \, \omega_2^{\Delta_2} \cdots \omega_n^{\Delta_n} \\
 \nonumber & \quad\times\! \left( \left[\frac{z_{n1}}{z_{n2}}\frac{\partial}{\partial\omega_2}+ \frac{z_{12}}{z_{n2}}\frac{\partial}{\partial\omega_n}\right] + \left[\frac{z_{n1}\bar{z}_{12}}{z_{n2}\omega_2} \frac{\partial}{\partial \bar{z}_2} + \frac{z_{12}\bar{z}_{1n}}{z_{n2}\omega_2} \frac{\partial}{\partial \bar{z}_n}\right]+ \left[ \frac{z_{n1}}{z_{n2}}\frac{2s_2}{\omega_2} + \frac{z_{12}}{z_{n2}}\frac{2s_n}{\omega_n} \right]\right)^{1-k}\!\!\!\mA_{n-1}\\
 &\nonumber = \operatorname{Tr}\,[T^{a_1} T^{a_2} \dots T^{a_n} ] \sum_{\mb=0}^{1-k} H^{k,+1,a_1}_{-\mb-\frac{k-1}{2}}(z_1)\zb_1^{\mb}\tilde{\mA}_{n-1},
\end{align*}
which is a simple polynomial of degree $1-k$ in $\zb_1$. A similar analysis has already been performed for gravity and can be found in~\cite{guevara2019notes}.
\subsection{NMHV and beyond}
The factorization of the amplitude in (\ref{eq:MHVfactorization}) ceases to be valid beyond the MHV sector.\footnote{Recently, an infinite number of soft theorems involving a part of the amplitude have been found in~\cite{Hamada:2018vrw, Li:2018gnc}.} While the leading and subleading soft limits are universal~\cite{Casali:2014xpa, Weinberg:1965nx}, the remaining terms are specific to the amplitude in question. We examine the series expansion of the NMHV 6-point amplitude and show that the amplitude is no longer a finite polynomial in the $z$, $\zb$ coordinates of the soft particle. This suggests that the polynomial structure of soft current breaks down in the presence of more than two negative helicity gluons. Consider the following expression for the 6-point NMHV (see for, e.g.,~\cite{Elvang:2013cua})
\begin{align*}
%\label{eq:6pmtNMHVorder0}
&\mA_6 \big(1^-, 2^-, 3^-, 4^+, 5^+, 6^+\big) \\
&\qquad= \frac{\lan{3}1+2\rsq{6}^3\, \delta^{4} (P)}{P^2_{126}\sq{21}\sq{16}\an{34}\an{45}\lan{5}1+6\rsq{2}}+\frac{\lan{1}5+6\rsq{4}^3\, \delta^{4} (P)}{P^2_{156}\sq{23}\sq{34}\an{56}\an{61}\lan{5}1+6\rsq{2}}, \nonumber
\end{align*}
where $\delta^{4} (P)$ is the momentum conserving delta function. If particle 1 becomes soft, we can expand the amplitude as series around $\omega_1 = 0$. The coefficient of $\omega_1^k$ can then be identified with the correlation function of the soft current as seen from (\ref{eq:softcurrentcorrelator}). The structure of these correlation functions is displayed below:
\begin{align}
&\big\langle H^{k,-1,a_1} (z_1, \zb_1) \mo^{-,a_2}_{\Delta_2}(z_2, \zb_2) \cdots \mo^{+,a_6}_{\Delta_6} (z_6, \zb_6)\big\rangle \nonumber\\
&\qquad =
 \begin{cases}
\displaystyle \frac{1}{(\zb_1-\zb_2)(\zb_1-\zb_6)}a_{0,0} , & k=1,\\
\displaystyle \frac{1}{(\zb_1-\zb_2)(\zb_1-\zb_6)} \sum_{i=0,j=0}^{i=1,j=1} a_{i,j}z_1^i \zb^j_1 , & k=0, \\
\displaystyle \frac{1}{(\zb_1-\zb_2)(\zb_1-\zb_6)}\frac{1}{(z_1-z_6)} \sum_{i=0,j=0}^{i=3,j=2} a_{i,j}z_1^i \zb^j_1 , & k=-1, \\
\displaystyle \frac{1}{(\zb_1-\zb_2)(\zb_1-\zb_6)}\frac{1}{(z_1-z_6)} \sum_{i=0,j=0}^{i=4,j=3} a_{i,j}z_1^i \zb^j_1 , & k=-2.
 \end{cases} \label{eq:NMHVsoftcurrents1}
\end{align}
Repeating the same exercise for particle 2 yields, we have
\begin{align}
 \label{eq:NMHVsoftcurrents2}
&\big\langle\mo^{-,a_1}_{\Delta_1}(z_1, \zb_1) H^{k,-1,a_2} (z_2, \zb_2) \cdots \mo^{+,a_6}_{\Delta_6} (z_6, \zb_6)\big\rangle \nonumber\\
 &\qquad=
 \begin{cases}
\displaystyle \frac{1}{(\zb_2-\zb_1)(\zb_2-\zb_3)}a_{0,0} , & k=1,\\
\displaystyle \frac{1}{(\zb_2-\zb_1)(\zb_2-\zb_3)}\sum_{i=0,j=0}^{i=1,j=1} a_{i,j}z_1^i \zb^j_1 , & k=0, \\
\displaystyle \frac{1}{(\zb_2-\zb_1)(\zb_2-\zb_3)}\sum_{i=0,j=0}^{i=2,j=2} a_{i,j}z_1^i \zb^j_1 , & k=-1, \\
\displaystyle \frac{1}{(\zb_2-\zb_1)(\zb_2-\zb_3)} \sum_{i=0,j=0}^{i=3,j=3} a_{i,j}z_1^i \zb^j_1 , & k=-2,
 \end{cases}
\end{align}
where we have omitted writing the overall trace factor. The leading and subleading soft factors are universal as expected. However, the soft currents starting from $k=-1$ develop a pole in~$z$ as seen in (\ref{eq:NMHVsoftcurrents1}). This is the pole which gives the collinear limit $1^-||6^+$. The full amplitude is a~sum over all orderings and will consequently have poles in both $z$ and $\zb$.

\subsection{Single soft factors in non-minimally coupled theories}
We examine the structure of the soft currents when we include
non-minimal couplings. In particular, we consider the example of Yang--Mills theory supplemented by an $F^3$ interaction. First order corrections to amplitude due to the $F^3$ operator have been computed in~\cite{Dixon:2004za}. The 6-point NMHV amplitude, including the first order corrections is
\begin{align*}
 %\label{eq:6pmtNMHVorder1}
 & \mA_6 \big(1^+, 2^+, 3^+, 4^-, 5^-, 6^-\big) \\
 &\qquad=\frac{[ 3|1+2|6\rangle^3}{P^2_{126}\an{21}\an{16}\sq{34}\sq{45}[5|1+6|2\rangle}+\frac{[1|5+6|4\rangle^3}{P^2_{156}\an{23}\an{34}\sq{56}\sq{61}[5|1+6|2\rangle}\nonumber\\
 &\qquad\quad+\frac{\kappa_{F^3}}{2}\frac{\an{45}^2\an{56}^2\an{64}^2}{\an{12}\an{23}\an{34}\an{45}\an{56}\an{61}} + \frac{\kappa_{F^3}}{2} \frac{\sq{12}^2\sq{23}^2\sq{31}^2}{\sq{12}\sq{23}\sq{34}\sq{45}\sq{56}\sq{61}}. \nonumber
\end{align*}
Processing as in the previous sections, we obtain results similar to (\ref{eq:NMHVsoftcurrents1}) and (\ref{eq:NMHVsoftcurrents2}) with coefficients~$a_{i,j}$ receiving first order corrections. It is worth pointing out here that taking the collinear limit~$1||2$ here will not produce only the $p=1$ term in the first line of (\ref{eq:softcurrentOPEcolor}). The $p=2$ term in the first line would arise only once we include second order corrections. The remaining terms would arise only if we include the contributions from the scalars.

Equations (\ref{eq:NMHVsoftcurrents1}, \ref{eq:NMHVsoftcurrents2}) show that a general Ansatz for the soft current cannot be a finite polynomial in $z$, $\zb$ (which holds only in the MHV sector as in~\cite{guevara2021holographic}) and we must have
\begin{align*}
 %\label{eq:softansatz}
 H^{k,s,a} = \sum_{\mb=-\infty}^{-\hb} \sum_{m=-\infty}^{-h} H_{m,\mb}^{k,s,a}\frac{1}{z^{m+h}}\frac{1}{\zb^{\mb+\hb}}.
\end{align*}
For the purposes of this paper, we choose to explicitly write the mode expansion in $\zb$ with the modes being a function of $z$ as
\begin{align*}
 %\label{eq:zbarexp}
 H^{k,s,a} = \sum_{\mb=-\infty}^{-\hb} H_{\mb}^{k,s,a}(z)\frac{1}{\zb^{\mb+\hb}}.
\end{align*}

\section{Gluon OPEs from double soft limits}
\label{app:doublesoft}
In this section, we derive the soft current OPEs directly from {\it simultaneous} double soft limits of scattering amplitudes. We are mainly interested in the singular terms in the OPE and these correspond to
taking holomorphic and anti-holomorphic collinear limits on the simultaneous double soft limit factor of $n$-gluon amplitudes. We can expand this amplitude in a basis of color-ordered amplitudes as in equation (\ref{eq:colorordered}). Collinear poles $\frac{1}{z_{12}}$ or $\frac{1}{\zb_{12}}$ can only occur in the color-ordered amplitudes when particles 1 and 2 are adjacent. We will restrict our attention to the color-ordered amplitude with the canonical ordering. A similar result holds for the remaining relevant orderings.

The simultaneous double soft limit of two positive helicity gluons has been derived in~\cite{klose2015double,volovich2015double}. Keeping only the singular term as $z_1 \to z_2$, we have
\begin{equation*}%\label{ppDSL}
\mAn \big(1^+,2^+,3, \dots, n\big) \xrightarrow{z_{12}\rightarrow0} \frac{\an{n2}}{\an{n1}\an{12}}\exp\left(\frac{\an{n1}}{\an{n2}}\tilde{\lambda}^{\dot{\alpha}}_1\partial_{\tilde{\lambda}^{\dot{\alpha}}_2}+\frac{\an{12}}{\an{n2}}\tilde{\lambda}^{\dot{\alpha}}_1\partial_{\tilde{\lambda}^{\dot{\alpha}}_n}\right)\mA_{n-1} \big(2^+, \dots, n\big).
\end{equation*}
We parametrize the spinors as in (\ref{eq:vars}).
% \begin{equation}\label{param}
% \lambda_1=\varepsilon_i\sqrt{\omega_i}\, t_i\begin{pmatrix}1\\ z_1\end{pmatrix},\quad \tilde{\lambda}_i=\sqrt{\omega_i}\,t_i^{-1}\begin{pmatrix}1\\ \bar{z}_i\end{pmatrix}.
% \end{equation}
 In the collinear limit,
\begin{align}
&\omega_1=\tau\omega_p,\qquad \omega_2=(1-\tau)\,\omega_p,\qquad z_1 = z_2 = z_p \qquad \zb_p = \zb_2 + \tau \zb_{12}. \nonumber\\
&\tilde{\lambda}_2+\frac{\an{n1}}{\an{n2}}\tilde{\lambda}_1=\sqrt{\frac{\omega_p}{1-\tau}}\begin{pmatrix} 1\\ \bar{z}_p\end{pmatrix} = \frac{1}{\sqrt{1-\tau}} \tilde{\lambda}_p, \label{eq:holomorphicCollinear}
\end{align}
which then gives
\begin{align*}
\mAn (1^+,2^+,3, \dots, n) \xrightarrow{z_{12}\rightarrow 0}{}&\frac{1}{z_{12}}\frac{1}{\tau\, \omega_p} \mA_{n-1} \left( \left\lbrace \sqrt{1-\tau}\lambda_p, \frac{1}{\sqrt{1-\tau}}\tilde{\lambda}_p\right\rbrace, \dots, \big\lbrace \lambda_n, \tilde{\lambda}_n\big\rbrace \right)\\
\nonumber &= \frac{1}{z_{12}}\frac{1}{\tau(1-\tau)\, \omega_p} \mA_{n-1} \big( \big\lbrace \lambda_p, \tilde{\lambda}_p\big\rbrace, \dots, \big\lbrace \lambda_n, \tilde{\lambda}_n\big\rbrace \big).
\end{align*}
Recalling that particle labelled by $p$ is soft, we can perform a soft expansion of the amplitude to get
\begin{gather}
\label{eq:dsoftexp}
 \mAn \big(1^+,2^+,3, \dots, n\big) \xrightarrow{z_{12}\rightarrow 0} \frac{1}{z_{12}}\frac{1}{\tau(1-\tau)} \sum_{m=-1}^\infty \omega^m_p H^{m,1}\left(z_2, \zb_2 + \tau \zb_{12}\right) \mA^{(m)},
\end{gather}
where $\mA^{(-1)} = \mA^{(0)} = \mA_{n-2} \left(3, \dots, n\right)$ are universal and the rest are specific to the amplitude. We can get the OPE by performing a Mellin transform. Recall that the celestial amplitude, which is the correlation function of primary gluon operators is
\begin{align*}
 \tilde{A_n} \big(1^+, 2^+, 3, \dots, n\big) &= \big\langle\mo^{+,a_1}_{\Delta_1}(z_1, \zb_1) \mo^{+,a_2}_{\Delta_2}(z_2, \zb_2) \cdots \mo^{s_n,a_n}_{\Delta_n} (z_n, \zb_n)\big\rangle\\
 &=\int \D\omega_1 \cdots \D\omega_n \,\omega_1^{\Delta_1-1} \cdots \omega_n^{\Delta_n-1} \mAn (1^+, 2^+, 3, \dots, n).\nonumber
\end{align*}
The double soft limit is
\begin{align*}
 %\label{eq:doublesoftdef}
 \lim_{\epsilon \to 0} \epsilon^2 \big\langle\mo^{+,a_1}_{k+\epsilon}(z_1, \zb_1) \mo^{+,a_2}_{l+\epsilon}(z_2, \zb_2) \cdots \mo^{s_n,a_n}_{\Delta_n} (z_n, \zb_n)\big\rangle, \qquad k, l = 1, 0, -1, \dots.
\end{align*}
Plugging in the double soft expansion from (\ref{eq:dsoftexp}) and using the identity (\ref{deltaidentity}),
we get
\begin{gather*}
H^{k_1,+1,a_1}(z_1,\zb_1)H^{k_2,+1,a_2}(z_2,\zb_2)
\sim -\frac{{\rm i} f^{a_1a_2b}}{z_{12}}\sum^{\infty}_{n=0}\begin{pmatrix}2-k_1-k_2-n\\1-k_2\end{pmatrix}\frac{\bar{z}_{12}^n}{n!}\bar{\partial}^n H^{k_1+k_2-1,+1,b}.
\end{gather*}

The simultaneous double soft limit in the mixed helicity case was also derived in~\cite{klose2015double,volovich2015double}. The singular terms in the collinear limit are given by
\begin{equation*}
\begin{aligned}
\mA\big(1^+,2^-,3,\dots ,n\big)\xrightarrow{1\rightarrow2}{}&\frac{\an{n2}}{\an{n1}\an{12}}\exp\left(\frac{\an{n1}}{\an{n2}}\tilde{\lambda}^{\dot{\alpha}}_1\partial_{\tilde{\lambda}^{\dot{\alpha}}_2}\right)\mA\big(2^-,\dots ,n\big)\\
&+\frac{\sq{13}}{\sq{12}\sq{23}}\exp\left(\frac{\sq{23}}{\sq{13}}\lambda^\alpha_2\partial_{\lambda^\alpha_1}\right)\mA\big(1^+,\dots ,n\big).
\end{aligned}
\end{equation*}

We treat the holomorphic and anti-holomorphic collinear limits separately. In the $z_{12}\rightarrow0$ case, we use (\ref{eq:holomorphicCollinear}) and find
\begin{equation*}
\begin{aligned}
\mAn \big(1^+,2^-,3, \dots, n\big) \xrightarrow{z_{12}\rightarrow 0}{}&\frac{1}{z_{12}}\frac{1}{\tau\, \omega_p} \mA_{n-1} \left( \left\lbrace \sqrt{1-\tau}\lambda_p, \frac{1}{\sqrt{1-\tau}}\tilde{\lambda}_p\right\rbrace, \dots, \big\lbrace \lambda_n, \tilde{\lambda}_n\big\rbrace \right)\\
& =\frac{1}{z_{12}}\frac{1-\tau}{\tau\, \omega_p} \mA_{n-1} \big( \big\lbrace \lambda_p, \tilde{\lambda}_p\big\rbrace, \dots, \big\lbrace \lambda_n, \tilde{\lambda}_n\big\rbrace \big).
\end{aligned}
\end{equation*}

The procedure for performing the Mellin transform is the same as in the $++$ case, we just need to account for the difference in powers of $1-\tau$ and that the soft particle $p$ is now of negative helicity. We then find
\begin{equation*}
\begin{aligned}
&\big\langle H^{k_1,+1,a_1}(z_1,\zb_1) H^{k_2,-1,a_2}(z_2,\zb_2)\cdots \calO^{s_n,a_n}_{\Delta_n}(z_n,\zb_n)\big\rangle \xrightarrow{z_{12}\rightarrow0}\\
&\qquad{}-{\rm i}f^{a_1a_2b}\frac{1}{z_{12}}\sum^{\infty}_{n=0}\begin{pmatrix}-k_1-k_2-n\\-1-k_2\end{pmatrix}\frac{\bar{z}_{12}^n}{n!}\bar{\partial}^n H^{k_1+k_2-1,-1,b}(z_2,\bar{z}_2)\tilde{\mA}^{(k_1+k_2-1)}.
\end{aligned}
\end{equation*}

To take the anti-holomorphic collinear limit we take the following parametrization
\begin{equation*}
\begin{aligned}
&\omega_1=\tau\omega_p,\qquad \omega_2=(1-\tau)\omega_p,\qquad \zb_1=\zb_2=\zb_p,\qquad z_p=z_2+\tau z_{12},\\
&\lambda_1+\frac{\sq{23}}{\sq{13}}\lambda_2=\sqrt{\frac{\omega_p}{\tau}}\begin{pmatrix}1\\z_p\end{pmatrix}=\frac{1}{\sqrt{\tau}}\lambda_p,
\end{aligned}
\end{equation*}
which gives
\begin{equation*}
\begin{aligned}
\mAn\big(1^+,2^-,3,\dots ,n\big)\xrightarrow{\zb_{12}\rightarrow0}{}&\frac{1}{\zb_{12}}\frac{1}{(1-\tau)\,\omega_p}\mA_{n-1}\left(\left\{\frac{1}{\sqrt{\tau}}\lambda_p,\sqrt{\tau}\tilde{\lambda}_p\right\},\dots ,\{\lambda_n,\tilde{\lambda}_n\}\right)\\
&=\frac{1}{\bar{z}_{12}}\frac{\tau}{(1-\tau)\omega_p}\mA_{n-1}\left(\{\lambda_p,\tilde{\lambda}_p\},\dots ,\{\lambda_n,\tilde{\lambda}_n\}\right).
\end{aligned}
\end{equation*}

Taking the Mellin transform, we find
\begin{equation*}
\begin{aligned}
&\big\langle H^{k,+1,a_1}(z_1,\zb_1) H^{l,-1,a_2}(z_2,\zb_2)\cdots \calO^{s_n,a_n}_{\Delta_n}(z_n,\zb_n)\big\rangle\xrightarrow{\zb_{12}\rightarrow0}\\
& \qquad -{\rm i}f^{a_1a_2b}\frac{1}{\zb_{12}}\sum^{\infty}_{n=0}\begin{pmatrix}-k-l-n\\1-l\end{pmatrix}\frac{z_{12}^n}{n!}\partial^n H^{k+l-1,+1,b}(z_2,\bar{z}_2)\tilde{\mA}^{(k+l-1)}.
\end{aligned}
\end{equation*}

The full OPE is then
\begin{equation*}
\begin{aligned}
H^{k,+1,a_1}(z_1,\zb_1) H^{l,-1,a_2}(z_2,\zb_2)
&\sim-\frac{{\rm i}f^{a_1a_2b}}{z_{12}}\sum^{\infty}_{n=0}\begin{pmatrix}-k-l-n\\-1-l\end{pmatrix}
\frac{\bar{z}_{12}^n}{n!}\bar{\partial}^n H^{k+l-1,-1,b}\\
& \quad -\frac{{\rm i}f^{a_1a_2b}}{\zb_{12}}\sum^\infty_{n=0}\begin{pmatrix}-k-l-n\\1-l\end{pmatrix}
\frac{z^n_{12}}{n!}\partial^n H^{k+l-1,+1,b}.
\end{aligned}
\end{equation*}

\section{Examples for deriving constraints of coupling constants}

In this appendix we show how we derive the constraint
\begin{equation}\label{eq:sampleconstraint}
 3\kappa_{0,2,2}^2 = 10 \kappa_{-2,2,2}\, \kappa_{2,2,2}
\end{equation}
in (\ref{eq:constraint1}). The other constraints can be derived similarly. Consider the double commutator (\ref{eq:posgravjoacobi}). Take $s_1=s_2=s_3=2$. We then have
\begin{align*}
 &\big[ \big[ W_{\mb_1}^{q_1,2}, W_{\mb_2}^{q_2,2} \big],W_{\mb_3}^{q_3,2}\big] + \text{cyclic} \\
&\qquad{} =\nonumber \sum_{l = 2,4,6} \sum_{\substack{r = 0\\ r \text{ odd}}}^l \frac{\kappa_{2, 2, l-r-3}}{2}
 \frac{\kappa_{3-l+r, 2, l-4}}{2} \\
 &\nonumber \qquad\qquad\qquad\quad \times N(q_1, q_2, \mb_1, \mb_2, l-r) N(q_1+q_2-l-1+r, q_3, \mb_1+\mb_2, \mb_3,r) \\
 & \qquad\qquad\qquad\quad \times W^{q_1+q_2+q_3-l-2, 4-l}_{\mb_1+\mb_2+\mb_3} \nonumber + \text{cyclic}.
\end{align*}
Here we are interested in the coefficients of the generator $W^{q_1+q_2+q_3-8,-2}$ on the right-hand side. These terms correspond to $l=6$, therefore $r$ can only be 1, 3, 5 in the summation. In the end the total coefficient is
\begin{gather}
 \sum_{r=1,3,5}\!\!\! \frac{\kappa_{2,2,3-r}}{2} \frac{\kappa_{r-3,2,2}}{2} N(q_1,q_2,\bar{m}_1,\bar{m}_2,6-r) N(q_1+q_2-7+r,q_3,\mb_1+\mb_2,\mb_3,r) \!+ \!\text{cyclic} \nonumber\\
\qquad = \frac{\kappa_{-2,2,2}\kappa_{2,2,2}}{4} \big( N(q_1,q_2,\mb_1,\mb_2,1) N(q_1+q_2-2,\mb_1+\mb_2,\mb_3,5) \nonumber\\
 \qquad \hphantom{\frac{\kappa_{-2,2,2}\kappa_{2,2,2}}{4} \Big(}{} + N(q_1,q_2,\mb_1,\mb_2,5) N(q_1+q_2-6,q_3,\mb_1+\mb_2,\mb_3,1) \big) \nonumber\\
 \qquad\quad{}+ \frac{\kappa_{0,2,2}^2}{4} N(q_1,q_2,\mb_1,\mb_2,3)\, N(q_1+q_2-4,q_3,\mb_1+\mb_2,\mb_3,3) + \text{cyclic}.
 \label{eq:3N}
\end{gather}
The $N$ polynomials satisfy the following identity:
\begin{equation*}
 \begin{aligned}
 & \frac{1}{5!} N(q_1,q_2,\mb_1,\mb_2,1)\, N(q_1+q_2-2,\mb_1+\mb_2,\mb_3,5) \\
 &\quad+ \frac{1}{5!} N(q_1,q_2,\mb_1,\mb_2,5)\, N(q_1+q_2-6,q_3,\mb_1+\mb_2,\mb_3,1)\\
 &\quad+ \frac{1}{3!^2} N(q_1,q_2,\mb_1,\mb_2,3)\, N(q_1+q_2-4,q_3,\mb_1+\mb_2,\mb_3,3) + \text{cyclic}\ =\ 0 .
 \end{aligned}
\end{equation*}
Therefore, (\ref{eq:3N}) becomes
\begin{gather*}
 \bigg( \frac{\kappa_{-2,2,2}\kappa_{2,2,2}}{4} - 3!^2 \, \frac{\kappa_{0,2,2}^2}{4 \times 5!} \bigg) \big( N(q_1,q_2,\mb_1,\mb_2,1) N(q_1+q_2-2,\mb_1+\mb_2,\mb_3,5) \\
 \qquad\qquad\qquad\quad{}+ N(q_1,q_2,\mb_1,\mb_2,5) N(q_1+q_2-6,q_3,\mb_1+\mb_2,\mb_3,1) \big) + \text{cyclic} = 0 .
 \end{gather*}
For arbitrary $q_1$, $q_2$, $\mb_1$, $\mb_2$, the quadratics of $N$ polynomials shown above are all independent, therefore to make (\ref{eq:3N}) vanishing, one get
\begin{equation*}
 \kappa_{-2,2,2}\kappa_{2,2,2} = \frac{3!^2}{5!} \kappa_{0,2,2}^2 = \frac{3 \kappa_{0,2,2}^2}{10},
\end{equation*}
which is exactly the constraint (\ref{eq:sampleconstraint}).

\section{Alternate definition of commutators}
\label{app:altcomm}
\begin{gather*}
% \label{eq:modeexpansion}
 H^{k, s}(z, \zb) =\overset{-\hb}{\underset{\mb = -\infty}{\sum}}\, \overset{-h}{\underset{m = -\infty}{\sum}} \frac{1}{\zb^{\mb+\hb}z^{m+h}} H_{m,\mb}^{k, s}, \\
 H^{k,s}_{m,\bar{m}} = \oint \frac{\D z}{2\pi {\rm i}} \oint \frac{\D \bar{z}}{2\pi {\rm i}} z^{m + h-1} \bar{z}^{\bar{m} + \hb-1} H^{k,s}(z , \bar{z}) .
\end{gather*}
Having defined this mode expansion, we can define the commutator
\begin{gather*}
 %\label{eq:badcommutator}
 \big[H_{m_1,\mb_1}^{k_1, s_1}, H_{m_2,\mb_2}^{k_2, s_2} \big] := \underset{z_1 > z_2}{\oint} \frac{{\rm d}z_1}{2\pi {\rm i}} \underset{z_2=0}{\oint}\frac{{\rm d}z_2}{2\pi {\rm i}} \underset{\zb_1 >\zb_2}{\oint}\frac{{\rm d}\zb_1}{2\pi {\rm i}} \underset{\zb_2 = 0}{\oint} \frac{{\rm d}\zb_2}{2\pi {\rm i}} H^{k_1, s_1} (z_1, \zb_1) H^{k_2, s_2} (z_2, \zb_2) \\
 \hphantom{\big[H_{m_1,\mb_1}^{k_1, s_1}, H_{m_2,\mb_2}^{k_2, s_2} \big] :=}{} -\underset{z_1 =0}{\oint} \frac{{\rm d}z_1}{2\pi {\rm i}} \underset{z_2> z_1}{\oint}\frac{{\rm d}z_2}{2\pi {\rm i}} \underset{\zb_1 = 0}{\oint}\frac{{\rm d}\zb_1}{2\pi {\rm i}} \underset{\zb_2 > \zb_1}{\oint} \frac{{\rm d}\zb_2}{2\pi {\rm i}} H^{k_2, s_2} (z_2, \zb_2) H^{k_1, s_1} (z_1, \zb_1) \\
 \hphantom{\big[H_{m_1,\mb_1}^{k_1, s_1}, H_{m_2,\mb_2}^{k_2, s_2} \big]}{} =\underset{z_1 = z_2}{\oint} \frac{{\rm d}z_1}{2\pi {\rm i}} \underset{z_2=0}{\oint}\frac{{\rm d}z_2}{2\pi {\rm i}} \underset{\zb_1 >\zb_2}{\oint}\frac{{\rm d}\zb_1}{2\pi {\rm i}} \underset{\zb_2 = 0}{\oint} \frac{{\rm d}\zb_2}{2\pi {\rm i}} H^{k_1, s_1} (z_1, \zb_1) H^{k_2, s_2} (z_2, \zb_2) \\
 \hphantom{\big[H_{m_1,\mb_1}^{k_1, s_1}, H_{m_2,\mb_2}^{k_2, s_2} \big] :=}{} + \underset{z_1 =0}{\oint} \frac{{\rm d}z_1}{2\pi {\rm i}} \underset{z_2> z_1}{\oint}\frac{{\rm d}z_2}{2\pi {\rm i}} \underset{\zb_1 = \zb_2}{\oint}\frac{{\rm d}\zb_1}{2\pi {\rm i}} \underset{\zb_2 =0}{\oint} \frac{{\rm d}\zb_2}{2\pi {\rm i}} H^{k_2, s_2} (z_2, \zb_2) H^{k_1, s_1} (z_1, \zb_1),
\\
 \big\lbrack H^{k_1, s_1}_{m_1, \bar{m}_1} ,\, H^{k_2, s_2}_{m_2, \bar{m}_2} \big\rbrack = -\sum_{p} \frac{\kappa_{s_1,s_2,-s_I}}{2}\\
 \qquad\times\sum_{x=0}^p
 \frac{ (\bar{m}_1 + \bar{m}_2 - \bar{h}_1 - \bar{h}_2 - p)! (-\bar{m}_1 - \bar{m}_2 - \bar{h}_1 - \bar{h}_2 - p)! }{(\bar{m}_1 - \bar{h}_1 - p + x)! (-\bar{m}_1 - \bar{h}_1 -x)! (\bar{m}_2 - \bar{h}_2 -x)! (-\bar{m}_2 - \bar{h}_2 - p +x)! }
 \\ \quad\qquad\qquad \times
 (-1)^{p-x} \frac{p!}{x! (p-x)!}
 H^{k_1 + k_2 + p - 1, s_1 + s_2 -p-1}_{m_1 + m_2 , \bar{m}_2 + \bar{m}_2} \\
 -\sum_{p} \frac{\bar{\kappa}_{s_1,s_2,-s_I}}{2} \!\sum_{x=0}^p
 \frac{ ({m}_1 + {m}_2 - {h}_1 - {h}_2 - p)! (-{m}_1 - {m}_2 - {h}_1 - {h}_2 - p)! }{({m}_1 - {h}_1 - p + x)! (-{m}_1 - {h}_1 -x)! ({m}_2 - {h}_2 -x)! (-{m}_2 - {h}_2 - p +x)! }
 \\ \qquad\qquad\qquad\qquad \quad\times
 (-1)^{p-x} \frac{p!}{x! (p-x)!}
 H^{k_1 + k_2 + p - 1, s_1 + s_2 +p+1}_{m_1 + m_2 , \bar{m}_2 + \bar{m}_2}.
 \end{gather*}
The Jacobi identity for commutators involving both positive and negative helicity currents is violated.

\subsection*{Acknowledgments}
We are grateful to Yangrui Hu, Luke Lippstreu, Marcus Spradlin and Andy Strominger for useful comments and discussion. This work was supported in part by the US Department of Energy under contract {DE}-{SC}0010010 Task A and by Simons Investigator Award~\#376208.

\pdfbookmark[1]{References}{ref}
\LastPageEnding

\end{document}